\documentclass[twocolumn,preprintnumbers,superscriptaddress]{revtex4}

\usepackage{palatino}
\usepackage{epsf}
\usepackage{amsmath,amssymb}
\usepackage{latexsym}
\usepackage{calc}
\usepackage{color}
\usepackage{graphicx}
\usepackage{float}
\usepackage[shortlabels]{enumitem}
\usepackage{CJK}

\DeclareMathAlphabet\mathbfcal{OMS}{cmsy}{b}{n}

\begin{document}

\begin{CJK*}{GB}{} % Use default fonts from CJK (see below)
\title{Ultrafast dynamics with the exact factorization}
\author{Federica Agostini$^\dagger$}
\affiliation{Universit\'e Paris-Saclay, CNRS, Institut de Chimie Physique UMR8000, 91405, Orsay, France. Email: federica.agostini@universite-paris-saclay.fr}
\author{E. K. U. Gross}
\affiliation{Fritz Haber Center for Molecular Dynamics, Institute of Chemistry, The Hebrew University of Jerusalem, Jerusalem 91904, Israel}

%\date{\today}
\begin{abstract}
The exact factorization of the time-dependent electron-nuclear wavefunction has been employed successfully in the field of quantum molecular dynamics simulations for interpreting and simulating light-induced ultrafast processes. In this work, we summarize the major developments leading to the formulation of a trajectory-based approach, derived from the exact factorization equations, capable of dealing with nonadiabatic electronic processes, and including spin-orbit coupling and the non-perturbative effect of an external time-dependent field. This trajectory-based quantum-classical approach has been dubbed coupled-trajectory mixed quantum-classical (CT-MQC) algorithm, whose performance is tested here to study the photo-dissociation dynamics of IBr.
\end{abstract}
\maketitle
\end{CJK*}

{\footnotesize
\noindent
$^\dagger$~federica.agostini@universite-paris-saclay.fr
}

\section{Introduction} 
The theoretical description of nonequilibrium processes at the microscopic scale poses continuous challenges in many fields, such as molecular and chemical physics, condensed matter physics, and theoretical chemistry. Theory needs to be able to describe purely quantum effects such as electronic transitions~\cite{Zewail_S1988,Curchod_CR2018,Barbatti_CR2018,Curchod_WIRES2019,Corni_C2019,Persico_JCP2104,Cederbaum_ARPC2004,Subotnik_ARPC2016,Curchod_PCCP2020}, (de)coherence~\cite{Fleming_N2007,Scholes_N2017,Rozzi_NC2013,Agostini_EPJB2018,Persico_CTC2014,Blumberger_NC2019}, interferences~\cite{Curchod_JCP2016,Persico_CPL1995}, energy relaxation~\cite{Olivucci_CR2017,Agostini_JCP2021_1,Agostini_JCTC2020_2}, or phase transitions~\cite{Fausti_Science2011,Nicoletti_AdvOpt2016}, that are often the result of complex interactions between electrons and nuclei on ultrafast time scales and of the non-perturbative effect of time-dependent external fields. While quantum mechanics is the key to unravel these processes, actual applications and studies relying on computational methods require the introduction of approximations. In the domain of quantum molecular dynamics, approximations can be intended in various way, (i) to reduce the complexity of the original problem via simplified models, (ii) to make the underlying equations of motion computationally tractable based on mathematical or physical observations, or (iii) to neglect some effects in favor of others depending on the situations. 

In the present work, we will provide examples on these three strategies, limiting our study to light-induced ultrafast phenomena in isolated molecular systems, and employing the formalism of the exact factorization of the time-dependent electron-nuclear wavefunction~\cite{EF_bookchapter_2020,Gross_PRL2010,Gross_JCP2012}. Using this theoretical framework, particular attention is devoted to present the procedures yielding simplified equations of motion, i.e., point (ii) above, that have made in recent years the exact factorization a suitable ``tool'' to perform quantum molecular dynamics simulations~\cite{Curchod_WIRES2019,Gross_TDDFTbook2018}. 

Introducing the factored form of a molecular -- electronic and nuclear -- wavefunction allows one to rewrite the time-dependent Schr\"odinger equation as coupled evolution equations for the electronic and the nuclear components of the full wavefunction. Once the problem is decomposed, the idea that has motivated and driven our efforts in the last years is to circumvent the exponential scaling of quantum mechanical equations of motion by using classical-like trajectories to mimic nuclear dynamics while maintaining a quantum viewpoint for electronic dynamics~\cite{Gross_MP2013,Gross_EPL2014,Gross_JCP2014,Gross_PRL2015}.

The exact factorization naturally lends itself for such a \textsl{quantum-classical} scheme because the original problem, i.e., the molecular time-dependent Schr\"odinger equation, is decomposed into a (single) electronic and nuclear part without invoking any approximations. Furthermore, nuclear dynamics appears to be fairly standard, in the sense that the nuclei evolve according to a (new) time-dependent Schr\"odinger equation where the effect of the electrons is accounted for via time-dependent vector~\cite{Agostini_JPCL2017,Curchod_EPJB2018} and scalar potentials~\cite{Gross_PRL2013,Gross_JCP2015}. Being explicitly dependent on time, the\-se potentials are able to describe effects due to electronic excited states and induced or driven by external fields~\cite{Gross_PRL2010,Schmidt_PRA2017,Maitra_PRL2015,Suzuki_PRA2014,Gross_PRL2017}. Aside from being a suitable framework for developing approximate numerical schemes, the exact factorization offers a framework to interpret, and perhaps even disentangle, complex effects: on one hand, nonadiabaticity and quantum decoherence are often strongly related to each other and affected by nuclear motion~\cite{Agostini_CTC2019,Maitra_JCTC2018,Tavernelli_EPJB2018,Min_JPCL2018,Gross_JPCL2017,Gross_PRL2015,Gross_JCTC2016}; on the other hand, nuclear quantum effects, such as interferences and branching in configuration space, can manifest themselves as consequence of the coupling to electronic dynamics, and compete with other similar effects, such as tunneling~\cite{Ciccotti_JPCA2020,Ciccotti_EPJB2018,Suzuki_PRA2016}.

While a comprehensive discussion on these points is beyond the scope of this work, we focus here on the possibility of capturing nonadiabatic effects and quantum decoherence during photo-induced ultrafast processes, by me\-ans of the quantum-classical scheme derived from the exact factorization. For completeness, we treat at the same level spin-allowed electronic nonadiabatic transitions, induced by nuclear motion, and spin-forbidden electronic transitions mediated by spin-orbit coupling~\cite{Agostini_PRL2020,Agostini_JCTC2020_1}. Following recent developments~\cite{Agostini_JCP2021_2}, we present as well how to explicitly account for external time-dependent fields (only the case of a continuous-wave (cw) laser will be discussed). The possibility of including quantum nuclear effects by adopting a quantum trajectory approach is also mentioned.

In order to present the exact factorization, together with its quantum-classical formulation, and to apply the theory to the study of a molecular process, we organize the paper as follows. In Section~\ref{sec: theory} we briefly recall the exact factorization, but we mainly devote this section to the derivation of the trajectory-based procedure, dubbed coupled-trajectory mixed quantum-classical (CT-MQC), ultimately leading to an actual algorithm for quantum-classical molecular dynamics simulations. Section~\ref{sec: IBr} focuses on a simple molecular application by presenting the performance of CT-MQC in describing the photo-disso\-cia\-tion of IBr, including spin-orbit coupling. A general assessment of the work done so far on the exact factorization and on future directions is presented in Section~\ref{sec: perspectives}. Our conclusions are stated in Section~\ref{sec: conclusions}.

\section{Excited-state dynamics with the exact factorization formalism}\label{sec: theory}
This section is devoted to the presentation of the exact factorization formalism to set the basis for the derivation of the CT-MQC algorithm.

We consider a system of interacting electrons and nuclei, including spin-orbit coupling (SOC) and an external time-dependent field to the molecular Hamiltonian. Therefore, the system is described by the Hamiltonian $\hat H(\mathbf x,\mathbf R,t) = \hat T_n(\mathbf R)+\hat H_{BO}(\mathbf x,\mathbf R)+\hat H_{SO}(\mathbf x,\mathbf R)+\hat V(\mathbf x,\mathbf R,t)$, where $\hat T_n$ is the nuclear kinetic energy, $\hat H_{BO}$, i.e., the Born-Oppenheimer Hamiltonian, contains the electronic kinetic energy, together with the electronic and nuclear potentials, $\hat H_{SO}$ is the SOC, $\hat V$ is the external time-dependent potential. Electronic position-spin variables are labelled as $\mathbf x = [\mathbf r_1,\boldsymbol\sigma_1], [\mathbf r_2,\boldsymbol\sigma_2],\ldots,[\mathbf r_{N_{el}},\boldsymbol\sigma_{N_{el}}]$, and nuclear positions are labelled as $\mathbf R=\mathbf R_1, \mathbf R_2,\ldots,\mathbf R_{N_n}$. Note that, the electronic operators $\hat{H}_{BO}$ and $\hat V$ are block-diagonal in spin space, and the particular form~\cite{Snijders_JCP1993,Tavernelli_JCP2015,Marian_WIREs2012} chosen for $\hat{H}_{SO}$ does not affect any of the results presented below.

The solution of the time-dependent Schr\"odinger equation (TDSE) $i\hbar\partial_t \Psi(\mathbf x,\mathbf R,t) = \hat H(\mathbf x,\mathbf R,t)\Psi(\mathbf x,\mathbf R,t)$ is factored as~\cite{Gross_PRL2010}
\begin{align}\label{eqn: EF}
\Psi(\mathbf x,\mathbf{R},t) = \chi(\mathbf R,t)\Phi_{\mathbf R}(\mathbf x,t)
\end{align}
with $\chi(\mathbf R,t)$ the nuclear wavefunction, and $\Phi_{\mathbf R}(\mathbf x,t)$ the electronic conditional factor that parametrically depends on $\mathbf R$. The partial normalization condition $\int d\mathbf x \left|\Phi_{\mathbf R}(\mathbf x,t)\right|^2 $ $=1$ $\forall\,\mathbf R,t$~\cite{Alonso_JCP2013,Gross_JCP2013}
%\begin{align}\label{eqn: PNC}
%\int d\mathbf x \left|\Phi_{\mathbf R}(\mathbf x,t)\right|^2 = 
%\sum_{\sigma_1}\sum_{\sigma_2}\ldots \int d\mathbf r_{1}\int d\mathbf r_{2}\ldots \left|\Phi_{\mathbf R}(\mathbf r_1,\sigma_1, \mathbf r_2, \sigma_2, \ldots,t)\right|^2 = 
%1 \quad\forall \,\, \mathbf R,t
%\end{align}
guarantees that $|\chi(\mathbf R,t)|^2$ reproduces at all times the (exact) nuclear density obtained from $\Psi(\mathbf x,\mathbf{R},t)$. Here, the integral over $\mathbf x$ stands for an integral over the $3N_{el}$-dimensional electronic configuration space and $N_{el}$ sums over electronic spins. The evolution of $\chi(\mathbf R,t)$ and $\Phi_{\mathbf R}(\mathbf x,t)$ is given by
\begin{widetext}
\begin{align}
i\hbar \partial_t \chi(\mathbf R,t) =&\Bigg[\sum_{\nu=1}^{N_n}\frac{[-i\hbar\nabla_\nu + \mathbf A_\nu(\mathbf R,t)]^2}{2M_\nu} + \epsilon(\mathbf R,t)+v(\mathbf R,t)\Bigg] \chi(\mathbf R,t)\label{eqn: EF n}\\
i\hbar \partial_t  \Phi_{\mathbf R}(\mathbf x,t) =& \Big[\hat H_{BO}(\mathbf x,\mathbf R)+\hat H_{SO}(\mathbf x,\mathbf R)+\hat V(\mathbf x,\mathbf R,t)+\hat U_{en}\left[\Phi_{\mathbf R},\chi\right]-\epsilon(\mathbf R,t)-v(\mathbf R,t)\Big]\Phi_{\mathbf R}(\mathbf x,t)\label{eqn: EF el}
\end{align}
\end{widetext}
The coupled nuclear and electronic evolution equations are derived by inserting the product form~(\ref{eqn: EF}) of the molecular wavefunction into the TDSE and by using the partial normalization condition, as described in detail in Refs.~\cite{Gross_JCP2012,Alonso_JCP2013,Gross_JCP2013,Ciccotti_EPJB2018}. %Here and in the following we use the shorthand notation
%\begin{align}
%\frac{1}{2}\mathbf M^{-1}\big(-i\hbar\boldsymbol{\nabla} + \mathbf A(\mathbf R,t)\big)^2 = \sum_{\nu=1}^{N_n}\frac{[-i\hbar\nabla_\nu+\mathbf A_\nu(\mathbf R,t)]^2}{2M_\nu}
%\end{align}
In Eq.~(\ref{eqn: EF n}) the symbol $\nabla_\nu$ indicates the gradient taken with respect to the positions of the $\nu$-th nucleus and $M_\nu$ are the nuclear masses. The time-de\-pen\-dent vector potential is a three-dimensional vector for each value of the index $\nu$, and it is defined as~\cite{Requist_PRA2015,Requist_PRA2017,Agostini_JPCL2017,Curchod_EPJB2018}
\begin{align}\label{eqn: TDVP}
\mathbf A_\nu(\mathbf R,t) = \left\langle\Phi_{\mathbf R}(t)\right| \left.-i\hbar\nabla_\nu\Phi_{\mathbf R}(t)\right\rangle_{\mathbf x}
\end{align}
The time-dependent scalar potentials are
\begin{widetext}
\begin{align}
\epsilon(\mathbf R,t)=& \Big\langle\Phi_{\mathbf R}(t)\Big|\hat H_{BO}(\mathbf x,\mathbf R)+\hat H_{SO}(\mathbf x,\mathbf R)+\hat U_{en}\left[\Phi_{\mathbf R},\chi\right]-i\hbar \partial_t\Big|\Phi_{\mathbf R}(t)\Big\rangle_{\mathbf x}\label{eqn: TDPES}\\
v(\mathbf R,t) =& \left\langle\Phi_{\mathbf R}(t)\right|\hat V(\mathbf x,\mathbf R,t)\left|\Phi_{\mathbf R}(t)\right\rangle_{\mathbf x}\label{eqn: v TDPES}
\end{align}
\end{widetext}
and mediate the coupling between electrons and nuclei, beyond the adiabatic regime. We distinguish two scalar potential contributions in order to separate the effect of the external time-dependent field. The first contribution, given by Eq.~(\ref{eqn: TDPES}), will be referred to as time-dependent potential energy surface (TDPES)~\cite{Gross_PRL2013,Gross_MP2013,Min_PRL2014,Gross_JCP2015,Curchod_JCP2016,Maitra_PRL2019}; the second contribution, Eq.~(\ref{eqn: v TDPES}), is associated to the external field. The integration over $\mathbf x$ is indicated as $\langle \,\cdot \,\rangle_{\mathbf x}$. Like the vector potential and the TDPES, the external contribution, Eq.~(\ref{eqn: v TDPES}), is, in general, an N-body operator, even though the external potential $\hat V(\mathbf x,\mathbf R,t)$ (typically representing a laser pulse) is a one-body operator. 
The electron-nuclear coupling operator $\hat U_{en}\left[\Phi_{\mathbf R},\chi\right]$ is~\cite{Gross_EPL2014,Gross_JCP2014,Agostini_ADP2015,AgostiniEich_JCP2016}
\begin{widetext}
\begin{align}
&\hat U_{en} [\Phi_{\mathbf R},\chi]= \sum_{\nu=1}^{N_n}\Bigg(\frac{[-i\hbar \nabla_\nu-\mathbf A_\nu(\mathbf R,t)]^2}{2M_\nu}+\frac{1}{M_\nu}\left(\frac{-i\hbar\nabla_\nu\chi(\mathbf R,t)}{\chi(\mathbf R,t)}+\mathbf A_\nu(\mathbf R,t)\right)\big(-i\hbar\nabla_\nu-\mathbf A_\nu(\mathbf R,t)\big)\Bigg)\label{eqn: enco}
\end{align}
\end{widetext}
and depends explicitly on the nuclear wavefunction $\chi(\mathbf R,t)$, and implicitly on the electronic factor $\Phi_{\mathbf R}(\mathbf x,t)$, via its dependence on the vector potential, given by Eq.~(\ref{eqn: TDVP}).

The product form~(\ref{eqn: EF}) of the molecular wavefunction is invariant under a gauge-like $\mathbf R,t$-dependent phase transformation of the electronic and nuclear components. Fixing the gauge means to choose an expression for such a phase or imposing a condition on the ``gauge fields'' that indirectly defines the phase. The gauge fields are the vector potential and the TDPES, since they transform as standard gauge potentials if the electronic and the nuclear wavefunctions are modified by the gauge phase. Solving Eqs.~(\ref{eqn: EF n}) and~(\ref{eqn: EF el}) with any given choice of gauge leads to a unique solution of the TDSE, as expected.

In the next sections we describe the procedure and approximations ultimately leading to the CT-MQC equations. In Section~\ref{sec: nucl} we rewrite the nuclear TDSE~(\ref{eqn: EF n}) employing the so-called quantum hydrodynamic formalism~\cite{wyattbook} and devise a trajectory-based scheme to solve it by invoking a classical limit (on the nuclear degrees of freedom). In Section~\ref{sec: el}, we introduce the expansion of the electronic wavefunction on a basis and we show how to solve the electronic evolution equation~(\ref{eqn: EF el}) along the classical nuclear trajectories. In Section~\ref{sec: ctmqc} we summarize the algorithm and briefly discuss the computational procedure for its implementation. Note that, recently, an attempt to solve numerically exactly Eqs.~(\ref{eqn: EF n}) and~(\ref{eqn: EF el}) has been made~\cite{Gossel_JCP2019}, but numerical instabilities seem to develop at the early stages of the propagation even for a simple one-dimensional model. This study confirms the need for the development of approximate schemes, as the one described below.

\subsection{Solution based on characteristics of the nuclear equation}\label{sec: nucl}
The polar representation of the (complex-valued) nuclear wavefunction, $\chi(\mathbf R,t) = |\chi(\mathbf R,t)|\exp{[(i/\hbar)S(\mathbf R,t)]}$, when inserted into the nuclear TDSE~(\ref{eqn: EF n}), yields the two coupled (real) equations~\cite{Ciccotti_EPJB2018}
\begin{widetext}
\begin{align}
-\partial_t S(\mathbf R,t) =&\sum_{\nu=1}^{N_n}\frac{\left[\nabla_\nu S(\mathbf R,t)+\mathbf A_\nu(\mathbf R,t)\right]^2}{2M_\nu}+\epsilon(\mathbf R,t)+v(\mathbf R,t)+\mathcal Q(\mathbf R,t)\label{eqn: HJ}\\
-\partial_t\left|\chi(\mathbf R,t)\right|^2=&\sum_{\nu=1}^{N_n}\left[ \nabla_\nu\cdot \frac{\nabla_\nu S(\mathbf R,t)+\mathbf A_\nu(\mathbf R,t)}{M_\nu}\left|\chi(\mathbf R,t)\right|^2\right]\label{eqn: continuity}
\end{align}
\end{widetext}
The evolution equation for the phase $S(\mathbf R,t)$, Eq.~(\ref{eqn: HJ}), has a Hamilton-Jacobi form, where the Hamiltonian on the right-hand side contains a kinetic term, with $\nabla_\nu S(\mathbf R,t)+\mathbf A_\nu(\mathbf R,t)$ being the momentum field, two ``classical'' time-dependent potential terms, $\epsilon(\mathbf R,t)$ and $v(\mathbf R,t)$, and a quantum potential term~\cite{Lopreore1999,wyattbook}, 
\begin{align}
\mathcal Q(\mathbf R,t)=\sum_{\nu=1}^{N_n}\frac{-\hbar^2}{2M_\nu}\frac{\nabla_\nu^2|\chi(\mathbf R,t)|}{|\chi(\mathbf R,t)|}
\end{align}
Since the quantum potential depends on the modulus of the nuclear wavefunction, it couples the Hamilton-Jacobi equation to the continuity equation, Eq.~(\ref{eqn: continuity}), which describes the evolution of the nuclear probability density $|\chi(\mathbf R,t)|^2$.

A quantum trajectory scheme to solve the partial differential equation (PDE) Eq.~(\ref{eqn: HJ}) has been derived in Ref.~\-\cite{Ciccotti_EPJB2018} and relies on the method of characteristics. The PDE is replaced by a set of ordinary differential equations (O\-D\-Es), i.e., the \textsl{characteristic equations}, for the ``variables'' $\mathbf R(s)$, $t(s)$, $S(s)$, $\tilde{\mathbf P}(s)= \boldsymbol\nabla S(s)$, $\partial_tS(s)$ that always satisfy the original PDE, as functions of the parameter $s$. We indicate here the $3N_n$-dimensional gradient of $S$ as $\boldsymbol\nabla S(s)=\lbrace\nabla_\nu S(s)\rbrace_{\nu=1,\ldots,N_n}$ to simplify the notation. We introduce the function $F(\mathbf R,t,S,\tilde{\mathbf P},S_t)= S_t  + H_n(\tilde{\mathbf P},\mathbf R,t)=0$, whose total differential is $dF = F_{\mathbf R} \cdot d\mathbf R + F_t dt + F_SdS+F_{\tilde{\mathbf P}}\cdot d\tilde{\mathbf P} + F_{S_t}dS_t$. The quantities $F_{\mathbf R}$, $F_t$, $F_S$, $F_{\tilde{\mathbf P}}$, and $F_{S_t}$ indicate the partial derivatives of the function $F$ with respect to its variables. If a set of ``initial conditions'' is chosen such that $F=0$, the (non-trivial) characteristic equations representing the evolution of the variables $(\mathbf R,t,S,\tilde{\mathbf P},S_t)$ are
\begin{align}
\dot{\mathbf R}_\nu &= \frac{\tilde{\mathbf P}_\nu+\mathbf A_\nu}{M_\nu}\label{eqn: characteristic 1}\\
%\frac{d t(s)}{ds} &= 1\label{eqn: characteristic 2}\\
%\frac{dS(s)}{ds} &= \tilde{\mathbf P}\cdot\boldsymbol{\nabla}_{\tilde{\mathbf P}}H_n(\tilde{\mathbf P},\mathbf R,t)-H_n(\tilde{\mathbf P},\mathbf R,t)=\mathcal L  \left(\mathbf R,\mathbf M^{-1}\tilde{\mathbf P},t\right) \label{eqn: characteristic 3}\\
\dot{\tilde{\mathbf P}}_\nu &=  -\nabla_\nu H_n(\tilde{\mathbf P},\mathbf R,t)\label{eqn: characteristic 4}
%\frac{d S_t(s)}{ds} &= -\partial_t H_n(\tilde{\mathbf P},\mathbf R,t)\label{eqn: characteristic 5}
\end{align}
with $\nu=1,\ldots,N_n$, where the Hamiltonian $H_n(\tilde{\mathbf P},\mathbf R,t)$ is the right-hand side of Eq.~(\ref{eqn: HJ}). The symbols $\dot{\mathbf R}_\nu$ and $\dot{\tilde{\mathbf P}}_\nu$ are intended as total derivatives with respect to the parameter $s$ for each value of the nuclear index $\nu$. The characteristic equation $\dot t = 1$ shows that the parameter $s$ can be identified as the physical time $t$. The characteristic equations for $S$ and $S_t$, not reported here, can be easily derived from $\dot{\tilde{\mathbf P}}_\nu$, since $\tilde{\mathbf P}_\nu = \nabla_\nu S$. Equations~(\ref{eqn: characteristic 1}) and~(\ref{eqn: characteristic 4}) guarantee that the vector $(d\mathbf R,dt,dS,d\tilde{\mathbf P},dS_t)$ is orthogonal to the gradient $(F_{\mathbf R}, F_t,F_S,F_{\tilde{\mathbf P}},F_{S_t})$ of F, and thus along the characteristics $dF=0$. If at any time $t$ the characteristic ODEs are solved for any initial condition, the field $S(\mathbf R,t)$ (and $\nabla_\nu S(\mathbf R,t)$ as well) can be reconstructed.

We proposed different procedures to solve Eqs.~(\ref{eqn: HJ}) and~(\ref{eqn: continuity}): 
\begin{enumerate}[(i)]
\item The \textit{classical procedure}~\cite{Gross_PRL2015,Gross_JCTC2016,Gross_JPCL2017,Tavernelli_EPJB2018,Gross_TDDFTbook2018} relies on neglecting the quantum potential $\mathcal Q(\mathbf R,t)$, such that the equations decouple, in the sense that the equation for $S(\mathbf R,t)$ can be solved independently from the continuity equation. The nuclear density is reconstructed simply from the distribution of trajectories. Quantum effects such as tunnelling cannot be captured in this case.
\item The \textit{pseudo-quantum procedure}~\cite{Ciccotti_EPJB2018,Agostini_ADP2015} accounts for the quantum potential $\mathcal Q(\mathbf R,t)$ in the evolution for $S(\mathbf R,t)$. However, the nuclear density is only approximated as a sum of Gaussians centered at the position of the trajectories, without solving the continuity equation~(\ref{eqn: continuity}). Tunneling can be captured, even though the fine details of the nuclear density, and thus of the quantum potential, cannot be correctly reproduced, posing issues to capture effects such as interferences.
\item The \textit{quantum procedure}~\cite{Ciccotti_JPCA2020,Tronci_arXiv2020,Franco_JCP2017} is devised to solve the coupled equations for the phase $S(\mathbf R,t)$ and for the density $|\chi(\mathbf R,t)|^2$ according to Eqs.~(\ref{eqn: HJ}) and~(\ref{eqn: continuity}). This procedure is fully equivalent to the solution of the nuclear TDSE. However, so far, only a proof-of-principle study has been conducted in this direction~\cite{Ciccotti_JPCA2020}, due to the evident numerical instabilities caused by the quantum potential.
\end{enumerate}
Henceforth, all quantities depending on $\mathbf R$ become functions of $\mathbf R(t)$ as we solve the dynamics \textsl{along the flow} of characteristics. This is independent of the procedure chosen among the three possibilities just presented. Note that the term ``trajectory'' is used to indicate the collection of $3N_n$ nuclear coordinates that evolve in time.

The derivation of CT-MQC follows the classical procedure, thus $\mathcal Q(\mathbf R(t),t)=0$. In addition, after having calculated the gradient of the Hamiltonian on the right-hand side of Eq.~(\ref{eqn: characteristic 4}), we impose the condition $\epsilon(\mathbf R(t),t)+v\big(\mathbf R(t),t\big)+\sum_\nu\dot{\mathbf R}_\nu(t)\cdot\mathbf A_\nu(\mathbf R(t),t)=0$ to fix the gauge freedom. The characteristic equations thus reduce to 
\begin{align}
\dot{\mathbf R}_\nu(t) &= \frac{\mathbf P_\nu(t)}{M_\nu} \label{eqn: R dot}\\
\dot{\mathbf P}_\nu(t) &=\dot{\mathbf A}_\nu\big(\mathbf R(t),t\big)\label{eqn: P dot}
\end{align}
where we introduced the new symbol $\mathbf P_\nu = \tilde{\mathbf P}_\nu + \mathbf A_\nu$ for the classical momentum. Equation~(\ref{eqn: P dot}) shows that the force to be used in CT-MQC to propagate classical nuclear trajectories is derived from the time-dependent vector potential. Note that the SOC contribution to the TDPES, i.e., the second term in the definition given in Eq.~(\ref{eqn: TDPES}), has been gauged away and does not appear in Eq.~(\ref{eqn: P dot}).

In Section~\ref{sec: ctmqc} we provide a more explicit expression of the classical force, using the dependence of $\mathbf A_\nu(\mathbf R(t),t)$ on the electronic wavefunction. To this end, we describe in Section~\ref{sec: el} how to derive the evolution equation for $\Phi_{\mathbf R(t)}(\mathbf x,t)$ along the nuclear characteristics.

\subsection{Solution of the electronic equation along the characteristics}\label{sec: el}
Solving the electronic PDE~(\ref{eqn: EF el}) along the \textsl{flow} of nuclear trajectories -- the characteristics -- requires to switch from the Eulerian frame to the Lagrangian frame. In the Lagrangian frame, only total time derivatives can be evaluated \textsl{along the flow}, that is why the symbol $\dot{\Phi}_{\mathbf R(t)}(\mathbf x,t)$ has to be used, rather than $\partial_t \Phi_{\mathbf R(t)}(\mathbf x,t)$. This is done by applying the chain rule $\partial_t\Phi_{\mathbf R(t)}(\mathbf x,t)=\dot\Phi_{\mathbf R(t)}(\mathbf x,t)- \sum_\nu\dot{\mathbf R}_\nu(t)\cdot \nabla_\nu\Phi_{\mathbf R(t)}(\mathbf x,t)$ anywhere the partial time derivative is found.

The derivation of CT-MQC relies on the expansion of the electronic wavefunction $\Phi_{\mathbf R(t)}(\mathbf x,t)$ on a basis, namely
\begin{align}\label{eqn: el exp}
\Phi_{\mathbf R(t)}(\mathbf x,t) = \sum_m C_m\big(\mathbf R(t),t\big)\varphi_{\mathbf R(t)}^{(m)}(\mathbf x)
\end{align}
The electronic representation can be chosen in different ways, as illustrated here: 
\begin{enumerate}[(i)]
\item In the absence of an external time-dependent field and of SOC, the \textit{adiabatic representation} is usually preferred~\cite{Gross_JCTC2016}. The adiabatic basis is formed by the eigenstates of the Born-Oppenheimer Hamiltonian $\hat H_{BO}$. Transitions among electronic adiabatic states during the dynamics, that will emerge as a result of the coupled electron-nuclear motion, are induced by the nonadiabatic coupling (NAC). NAC is referred to as kinetic coupling, consequence of the action of the nuclear kinetic energy operator on the parametric dependence of the electronic states on the nuclear positions. Quantum chemistry softwares usually provide electronic structure information in the adiabatic basis, a feature that is crucial when combining CT-MQC with different approaches to electronic structure theory.
\item In the presence of SOC, the \textit{spin-diabatic} representation is often the chosen one~\cite{Agostini_JCTC2020_1,Curchod_JCP2016_SOC,Tavernelli_JCP2015}. Spin-diabatic states can be labeled depending on their spins, e.g., singlets, or triplets, and are, thus, easily employed for interpreting spectroscopic results. The spin-diabatic basis is formed by the eigenstates of the Born-Oppen\-heimer Hamiltonian $\hat H_{BO}$. Transitions among electronic spin-diabatic states of different spin-multi\-pli\-ci\-ty are induced by the SOC, whereas NAC can mediate transitions within the same spin multiplet. Alternatively, the \textit{spin-adiabatic} representation can be used, formed by the eigenstates of $\hat H_{BO}+\hat H_{SO}$. Spin-adiabatic states are combinations of different spin multiplicities and transition among them is purely of kinetic nature (NAC).
\item In the presence of an external time-dependent field, in particular of a laser pulse, a possible choice is the representation based on the eigenstates of the Hamiltonian $\hat H_{BO}$ (or $\hat H_{BO}+\hat H_{SO}$ if SOC is present), which has been called \textit{field-diabatic} representation in Ref.~\cite{Curchod_JPCA2019}. Transitions among electronic states are of NAC or SOC character, depending on the physical problem at hand. In addition, transitions can be induced by the field itself, respecting the spin selection rules, and are mediated by the transition dipole moment when only an external electric field is considered. 
\item In the case of a cw laser, the \textit{Floquet diabatic} representation~\cite{Shalashilin_CP2018,Schuette_JCP2001,Subotnik_JCTC2020,Schmidt_PRA2016,Gonzalez_JPCA2012,Welsch_PRA2020,Agostini_JCP2021_2,Bucksbaum_JPB2015} appears to be a viable option especially for interpreting absorption/emission processes~\cite{Mies_PRL1990,Taday_JPB2000,GiustiSuzor_PRA1988,Atabek_PRA1992,Schumacher_PRL1990,Langley_PRL2001,Schmidt_PRA2005,Schmidt_PRA2017} in terms of exchanges of photons between the system and the field. Floquet representation exploits the periodicity induced by the laser in the Hamiltonian~\cite{Sambe_PRA1973}, which is encoded in the mixed character of field-electronic states, defined as the eigenstates of $\hat H_{BO}-i\hbar\partial_t$. Transitions among Floquet diabatic states are either of NAC nature, conserving the number of photons, or are induced by photon exchanges. For completeness we mention as well the \textit{Floquet adiabatic} representation~\cite{Agostini_JCP2021_2}, formed by the eigenstates of $\hat H_{BO}+\hat V-i\hbar\partial_t$, even though its applications in the field of quantum molecular dynamics simulations have not been thoroughly investigated so far.
\end{enumerate}
In order to present a derivation of CT-MQC that is as general as possible we use here the spin-diabatic representation. Once the final equations are obtained, it is easy to see how they can be modified to accommodate alternative representations. In the following derivation, we do not consider the external field because including the effect of a cw laser via the Floquet formalism requires the use of a slightly more involved representation~\cite{Agostini_JCP2021_2}, whose implementation in the CT-MQC algorithm is still ongoing work.

The electronic CT-MQC equation describes the time evolution of the expansion coefficients introduced in Eq.~\-(\ref{eqn: el exp}). The purpose of the following discussion is, thus, to present a derivation of the quantity $\dot C_m\big(\mathbf R(t),t\big)$. However, we only describe the critical points of such derivation, since the details can be found in Refs.~\cite{Gross_JCTC2016,Gross_JPCL2017,Agostini_JCTC2020_1,Agostini_JCP2021_1}. 

In the definition~(\ref{eqn: enco}) of $\hat U_{en}$ we identify two terms: the first term has been shown~\cite{Scherrer_JCP2015,AgostiniEich_JCP2016,Scherrer_PRX2017} to be smaller if compared to the second term, and it is thus neglected. The corresponding term in the TDPES will be neglected as well to maintain gauge invariance of the exact factorization equations in their trajectory-based formulation~\cite{Gross_PRL2015,Gross_JCTC2016}. 
The electron-nuclear coupling operator depends on the nuclear wavefunction. When its polar representation is used, the expression of $\hat U_{en} $ reduces to
\begin{align}
\hat U_{en} &\simeq\sum_{\nu=1}^{N_n}\left(\dot{\mathbf R}_\nu(t)+i\frac{\boldsymbol{\mathcal P}_\nu\big(\mathbf R(t),t\big)}{M_\nu}\right)\left(-i\hbar\nabla_\nu-\mathbf A_\nu\big(\mathbf R(t),t\big)\right)\label{eqn: Uen with R(t) and Q(t)}
\end{align}
The first term in parenthesis of Eq.~(\ref{eqn: Uen with R(t) and Q(t)}) is the velocity of the trajectory from the characteristic equation~(\ref{eqn: R dot}), whereas the second term contains the quantum momentum~\cite{Gross_PRL2015}, $\boldsymbol{\mathcal P}_\nu(\mathbf R(t),t) = -\hbar|\chi(\mathbf R(t),t)|^{-1}\nabla_\nu|\chi(\mathbf R(t),t)|$, that has been defined in previous work on CT-MQC. The quantum momentum~\cite{Rassolov_JCP2004,Rassolov_CPL2003} induces quantum decoherence effects by tracking the spatial delocalization over time of the nuclear density (or, equivalently, of its modulus)~\cite{Agostini_EPJB2018,Maitra_JCTC2018}. Since the nuclear density at each time has to be reconstructed to evaluate its spatial derivative, and thus the quantum momentum, CT-MQC trajectories are ``coupled'', in the sense that they cannot be propagated independently from each other. CT-MQC equations encode some non-local information about the dynamics, which is the key to correctly capture the quantum mechanical effect of decoherence.

The expression of the electronic wavefunction along the trajectories in the chosen representation is inserted in Eq.~(\ref{eqn: el exp}). Note that when the total time derivative acts on the electronic states, only the gradient term survives, because they depend on time implicitly, via their dependence on the trajectory.

In order to isolate the evolution of one of the coefficients, say $ C_m\big(\mathbf R(t),t\big)$, one has to project the electronic equation onto $\bar\varphi_{\mathbf R(t)}^{(m)}(\mathbf x)$, by integrating over $\mathbf x$. The NAC vectors and the matrix elements of the SOC appear in the final equation, and are defined as $\mathbf d_{\nu,ml}^{\alpha} = \langle \varphi_{\mathbf R^\alpha(t)}^{(m)}|\nabla_\nu\varphi_{\mathbf R^\alpha(t)}^{(l)}\rangle_{\mathbf x}$ and $[H_{SO}^\alpha]_{ml}=\langle \varphi_{\mathbf R^\alpha(t)}^{(m)}|\hat H_{SO}(\mathbf R^\alpha(t))|\varphi_{\mathbf R^\alpha(t)}^{(l)}\rangle_{\mathbf x}$, respectively. Note that we used the superscript $\alpha$ to label the dependence on the trajectory $\mathbf R^\alpha(t)$.

The action of the Born-Oppenheimer Hamiltonian on its eigenstates yields the eigenvalues $E^\alpha_m$, clearly depending on the position of the trajectory.

As consequence of the action of the gradient operator $\nabla_\nu$ in Eq.~(\ref{eqn: Uen with R(t) and Q(t)}) on the electronic wavefunction, the equation contains a term like $\nabla_\nu C_m^\alpha(t)$, which is approximated by neglecting the spatial dependence of the modulus of $C_m^\alpha(t)$ in favor of the spatial dependence of its phase. Therefore, $\nabla_\nu C_m^\alpha(t)\simeq (i/\hbar)\mathbf f_{\nu,m}^{\alpha}C^{\alpha}_{m}(t)$ and we use a simple expression for the gradient of the phase, namely $\mathbf f_{\nu,m}^{\alpha} = \int_0^t d\tau[-\nabla_\nu E^{\alpha}_m]$; this quantity is a spin-diabatic force accumulated over time along the trajectory $\alpha$.

With these definitions and approximations in mind, we can rewrite the electronic evolution equation given in Eq.~(\ref{eqn: EF el}) in the spin-diabatic basis along the trajectory $\mathbf R^\alpha(t)$ as
%\begin{subequations}\label{eqn: C dot}
\begin{widetext}
\begin{align}
\dot C_m^{\alpha}(t) =& \Big[-\frac{i}{\hbar}E_m^\alpha + \sum_{\nu=1}^{N_n}\frac{\boldsymbol{\mathcal P}_\nu^\alpha(t)}{\hbar M_\nu}\cdot\big(\mathbf f_{\nu,m}^\alpha-\mathbf A_\nu^\alpha(t)\big)\Big]C_m^\alpha(t)-\sum_{l}\left(\frac{i}{\hbar} \left[ H_{SO}^\alpha\right]_{ml}+\sum_\nu\dot{\mathbf R}_\nu^\alpha(t) \cdot \mathbf d_{\nu,ml}^\alpha\right)C_m^\alpha(t)\label{eqn: C dot}
%& -\frac{i}{\hbar}v^{\alpha}(t)C_m^\alpha(t)-\frac{i}{\hbar} \sum_{l}V^\alpha_{ml}(t)C_l^\alpha(t)\label{eqn: C dot 3}
\end{align} 
%\end{subequations}
\end{widetext}
Some quantities in this equation depend explicitly on time, and such dependence has been indicated together with the dependence on $\mathbf R^{\alpha}(t)$; some other quantities only depend on time via their dependence on the trajectory, which is indicated with the superscript $\alpha$.

The first diagonal term in Eq.~(\ref{eqn: C dot}) contains in square brackets a purely imaginary part, depending on the spin-diabatic energy, and a purely real part, depending on the quantum momentum. The former is responsible for the oscillating phase of $C_m^{\alpha}(t)$, the latter affects the modulus of $C_m^{\alpha}(t)$ and is, thus, source of decoherence effects. It is worth mentioning here that the norm of the electronic wavefunction is preserved along the dynamics by Eq.~(\ref{eqn: C dot}), i.e., $\sum_m|C_m^{\alpha}(t)|^2=1$ $\forall\,t,\alpha$. The off-diagonal contributions in Eq.~(\ref{eqn: C dot}) drive electronic transitions from state $m$ to state $l$, mediated by the SOC, if states with different spin multiplicity are concerned, and by the NAC vectors, if states with the spin multiplicity are involved.
%Last term, Eq.~(\ref{eqn: C dot 3}), contains the effects arising from the action of the external time-dependent field on the system, via a diagonal contribution, depending on the scalar time-dependent potential of Eq.~(\ref{eqn: v TDPES}), and an off-diagonal contribution, which induces electronic transitions mediated by the external field. The Floquet-based representation of the terms in Eq.~(\ref{eqn: C dot 3}) has been derived in Ref.~[\!\citenum{Agostini_JCP2021_2}].

The time-dependent vector potential appears in Eq.~(\ref{eqn: C dot}). According to its definition given in Eq.~(\ref{eqn: TDVP}), it depends on the electronic wavefunction, therefore, it can be expressed in the chosen electronic basis along a trajectory $\mathbf R^\alpha(t)$ as $\mathbf A_\nu^\alpha(t)=-i\hbar\sum_m\bar C_m^{\alpha}(t)\nabla_\nu C_m^{\alpha}(t)-i\hbar\sum_{m,l}\bar C_m^{\alpha}(t)C_l^{\alpha}(t)\mathbf d_{\nu,mj}^{\alpha}$. Recalling the approximation used above to express the gradient of the electronic coefficients, and observing that the first term of this expression accumulates over time, while the second is localized in space due to the dependence on the NAC vectors, the CT-MQC expression of the (real-valued) time-dependent vector potential is
\begin{align}
\mathbf A_\nu^\alpha(t)=\sum_m\left|C_m^{\alpha}(t)\right|^2\mathbf f_{\nu,m}^\alpha
\end{align} 
The classical force of Eq.~(\ref{eqn: P dot}) used to propagate CT-MQC trajectories contains the vector potential, and its total time derivative has to be computed to give an explicit expression of the force in terms of electronic coefficients and the other electronic properties introduced in Eq.~(\ref{eqn: C dot}). The final expression in given in Section~\ref{sec: ctmqc}.

\subsection{CT-MQC algorithm}\label{sec: ctmqc}
Sections~\ref{sec: nucl} and~\ref{sec: el} described the procedure and the approximations used to derive CT-MQC electronic and nuclear evolution equations. Essentially, those equations are a set of ODEs, thus referred to as \textsl{quantum-classical}, representing a way to reformulate the quantum mechanical equations of the exact factorization, i.e., Eqs.~(\ref{eqn: EF n}) and~(\ref{eqn: EF el}). In particular, the electronic equation~(\ref{eqn: C dot}) is solved along a trajectory, thus it lends itself for an on-the-fly approach, where electronic structure properties are only computed at the instantaneous nuclear positions. In turn, the nuclei evolve according to a classical-like force determined by the instantaneous state of the electrons. 

Nuclear forces of Eq.~(\ref{eqn: P dot}) are determined by computing the total time derivative of the vector potential. Writing explicitly this derivative yields the following expression of the force for the trajectory $\mathbf R^\alpha(t)$
\begin{align}\label{eqn: force}
\mathbf F_\nu^{\alpha}(t) = \mathbf F^{\alpha}_{\nu,\textrm{Eh}}(t)+ \mathbf F^{\alpha}_{\nu,\textrm{qm}}(t)+ \mathbf F^{\alpha}_{\nu,\textrm{SOC}}(t)%+ \mathbf F^{\alpha}_{\textrm{ext}}(t)
\end{align}
The Ehrenfest-like term (Eh) is a standard mean-field force
\begin{align}
\mathbf F^{\alpha}_{\nu,\textrm{Eh}}(t) =& \sum_{m} \left|C_{m}^{\alpha}(t)\right|^2 \left(-\nabla_\nu E_{m}^{\alpha}\right)\nonumber\\
&+\sum_{m,l}\bar C_m^{\alpha}(t)C_l^{\alpha}(t)\left(E_{m}^{\alpha}-E_{l}^{\alpha}\right)\mathbf d_{\nu,ml}
\end{align}
The quantum momentum term (qm) depends on $\boldsymbol{\mathcal P}_\nu^\alpha(t)$, namely
\begin{align}
\mathbf F^{\alpha}_{\nu,\textrm{qm}}(t) =\frac{2}{\hbar}\sum_m\left|C_m^{\alpha}(t)\right|^2\left[\sum_{\mu=1}^{N_n}\boldsymbol{\mathcal P}_\mu^\alpha(t)\cdot \mathbf f_{\mu,m}^{\alpha}\right]\left(\mathbf f_{\nu,m}^{\alpha}-\mathbf A_\nu^{\alpha}(t)\right)
\end{align}
and couples the trajectories through the presence of the quantum momentum, as described in Section~\ref{sec: el}. The SOC term contains the matrix elements of the spin-orbit Hamiltonian
\begin{align}
\mathbf F^{\alpha}_{\nu,\textrm{SOC}}(t)=\frac{1}{\hbar}\sum_{m,l}\mathrm{Im}\left(\bar C^{\alpha}_{m}(t)C^{\alpha}_{l}(t)[H_{SO}^\alpha]_{ml}\right) \left(\mathbf f^{\alpha}_{\nu,l} (t)-\mathbf f^{\alpha}_{\nu,m} (t)\right)\label{eqn: SOC force}
\end{align}
It is clear from this expression that it is crucial to correctly capture decoherence effects in the presence of SOC. In fact, usually, SOC is very delocalized in (nuclear) space and, in some situations, it is even considered  spatially constant~\cite{Casavecchia_PNAS2012}. This feature is clearly different from the behavior of NAC, which is typically localized in the regions of avoided crossings and conical intersections. If the SOC contribution to the force, Eq.~(\ref{eqn: SOC force}), goes to zero, it does so as effect of decoherence, and not because the SOC itself goes to zero. Decoherence manifests itself as the ``collapse'' of the electronic time-dependent wavefunction along a given trajectory to a single electronic state, with the corresponding coefficient becoming one. Norm conservation along that trajectory imposes that all other coefficients become zero, finally yielding no contribution from the SOC force as consequence of the decoherence process.

It should be noted that adapting standard trajectory-based algorithms for nonadiabatic dynamics to the delocalized nature of SOC requires to revisit those algorithms if calculations are performed in the spin-diabatic basis. In fact, methods such as ab initio multiple spawning~\cite{Curchod_JCP2016_SOC,Varganov_JPCA2018,Varganov_JPCA2016} and surface hopping~~\cite{Gonzalez_JCTC2011,Gonzalez_IJQC2015,Thiel_JCP2014,Tavernelli_JCP2015} rely on the fact that NACs are spatially localized. The choice between spin-diabatic or spin-adibatic basis when dealing with SOC depends on the quantum-chemistry package chosen for the electronic-structure calculations; quantum-chemistry codes typically provide electronic-stru\-ctu\-re information in the spin-diabatic basis. However, if calculations are possible in the spin-adiabatic basis, all couplings become NACs, thus spatially localized, and the algorithms can be employed in their original form. The interested reader is referred to Ref.~\cite{Agostini_JCTC2020_1} for a detailed discussion on this topic with connections to various trajectory-based approaches considering SOC.

%The force contribution arising from the presence of the external field is
%\begin{align}
%\mathbf F^{\alpha}_{\textrm{ext}}(t)=&-\sum_{m,l}\boldsymbol{\nabla}V_{lm}^\alpha(t)\bar C_{l}^\alpha(t)C_{m}^\alpha(t)+\frac{1}{\hbar}\sum_{m,l}\mathrm{Im}\left[\bar C_{m}^\alpha(t)C_{l}^\alpha(t)\right]V_{ml}^\alpha(t)\left(\mathbf f_{l}^\alpha-\mathbf f_{m}^\alpha\right)
%\end{align}
%Note that the second term on the right-hand side is formally identical with the SOC term, while the first term appears as consequence of the choice of the gauge. In fact, in fixing the gauge freedom, we have included the SOC contribution in the scalar TDPES that is gauged away but not the external field contribution.

Similarly to the expression of the force given in Eq.~(\ref{eqn: force}), we can rewrite the electronic evolution equation~(\ref{eqn: C dot}) as
\begin{align}\label{eqn: C dot ctmqc}
\dot C_m^{\alpha}(t) = \dot C_m^{\alpha}(t)\Big|_{\textrm{Eh}}+\dot C_m^{\alpha}(t)\Big|_{\textrm{qm}}+\dot C_m^{\alpha}(t)\Big|_{\textrm{SOC}} %+\dot C_m^{\alpha}(t)\Big|_{\textrm{ext}}
\end{align}
where we identify the following terms
\begin{subequations}\label{eqn: C dot bis}
\begin{align}
\dot C_m^{\alpha}(t)\Big|_{\textrm{Eh}} &= -\frac{i}{\hbar}E_m^\alpha C_m^\alpha(t)-\sum_l\sum_{\nu=1}^{N_n}\dot{\mathbf R}_\nu^\alpha(t) \cdot \mathbf d_{\nu,ml}^\alpha C_l^\alpha(t)\label{eqn: C dot bis 1}\\
\dot C_m^{\alpha}(t)\Big|_{\textrm{qm}} &= \sum_{\nu=1}^{N_n}\frac{\boldsymbol{\mathcal P}_\nu^\alpha(t)}{\hbar M_\nu}\cdot\big(\mathbf f_{\nu,m}^\alpha-\mathbf A_\nu^\alpha(t)\big)C_m^\alpha(t)\label{eqn: C dot bis 1}\\
\dot C_m^{\alpha}(t)\Big|_{\textrm{SOC}} &=-\frac{i}{\hbar} \sum_{l}[ H_{SO}^\alpha]_{ml}C_l^\alpha(t) \label{eqn: C dot bis 3}
%\dot C_m^{\alpha}(t)\Big|_{\textrm{ext}} &= -\frac{i}{\hbar}v^{\alpha}(t)C_m^\alpha(t)-\frac{i}{\hbar} \sum_{l}V^\alpha_{ml}(t)C_l^\alpha(t)\label{eqn: C dot bis 4}
\end{align}
\end{subequations}
In the absence of SOC, the term $[H_{SO}^\alpha]_{ml}$ in Eq.~(\ref{eqn: C dot bis 3}) is not present, and the electronic equation reduces to the one in the original derivation of CT-MQC in the adiabatic representation reported in Refs.~\cite{Gross_PRL2015}. In the presence of NAC and SOC, the algorithm is dubbed generalized CT-MQC (G-CT-MQC) in the spin-diabatic basis~\cite{Agostini_PRL2020}. As discussed earlier, in the presence of spin-orbit effects, one can transform the coupling mediated by the SOC into NAC in the spin-adiabatic representation. In this case, $[H_{SO}^\alpha]_{ml}$ of Eq.~(\ref{eqn: C dot bis 3}) does not appear and the electronic equation is identical with the original CT-MQC equation.

In the presence of a cw laser field, if the Floquet diabatic representation is used, the periodic time dependence induced by the external field is expressed in Fourier space, via the harmonics of the driving frequency. An additional term appears in the expression of the force of Eq.~(\ref{eqn: force}), namely, $\mathbf F^{\alpha}_{\nu,\textrm{ext}}(t)=\hbar^{-1}\sum_{m,l}\mathrm{Im}\left[\bar C_{m}^\alpha(t)C_{l}^\alpha(t)\right]V_{ml}^\alpha(\mathbf f_{\nu,l}^\alpha-$ $\mathbf f_{\nu,m}^\alpha)$, and in the expression of the evolution of the electronic coefficients of Eq.(\ref{eqn: C dot ctmqc}), namely $\dot C_m^{\alpha}(t)\big|_{\textrm{ext}}=-i\hbar^{-1} $ $ \sum_{l}V^\alpha_{ml}C_l^\alpha(t)$. Furthermore, the energies $E_m^\alpha$ of Eq.~(\ref{eqn: C dot bis 1}) are shifted by a fixed amount depending on the photon energy of the corresponding harmonic. Note that the indices $m,l$ in the Floquet picture label the electronic physical states as well as the harmonic of the driving frequency. Thus, $V^\alpha_{ml}$ induces transition between electronic states and between harmonics. We recall that the matrix elements of the external field in the Floquet diabatic basis are expressed in Fourier space (that is why $V^\alpha_{ml}$ does not depend on time). Using the Floquet diabatic representation, the algorithm is dubbed F-CT-MQC~\cite{Agostini_JCP2021_2}.

In the next section we apply the CT-MQC algorithm to study the photo-dissociation reaction of IBr. The molecule IBr is known to have strong spin-orbit effects, and thus, it is selected here to show the performance of (G-)CT-MQC in the spin-adiabatic and spin-diabatic flavors presented above.

\section{Photo-dissociation of IBr}\label{sec: IBr}
In this section we simulate the photo-induced dissociation of IBr based on CT-MQC and we focus on the calculation of the branching ratio of the products.

As reported in the literature~\cite{Leone_Science2019}, a strong transition dipole moment couples the ground-state of the molecule $\mathrm X(^1\Sigma_{0^+})$ to its excited electronic state $\mathrm B(^3\Pi_{0^+})$. In the Franck-Condon region, in particular, this transition dipole moment is the dominant one, thus we will consider that photo-dissociation is initiated by an excitation $\mathrm X(^1\Sigma_{0^+}) \rightarrow\mathrm B(^3\Pi_{0^+})$. The electronic states considered here are denoted using typical spin-diabatic labels~\cite{Gonzalez_JCTC2011,Worth_FD2011} even though we will work in the spin-adiabatic representation as well, based on the model potentials of Ref.~\cite{Guo_JCP1993}. In particular, the spin-diabatic characters $\mathrm B(^3\Pi_{0^+})$ and $\mathrm Y(0^+)$, discussed below, describe the first-excited and second-excited electronic states, respectively, in the Franck-Condon region $R<5$~bohr.

The molecule IBr manifests strong SOC, which is responsible for the appearance of an avoided crossing between the first-excited and second-excited electronic states, ultimately leading to the dissociation of the molecule via two channels: within 300~fs from photo-excitation, the products of such an ultrafast photo-reaction are $\mathrm I + \mathrm{Br}(^2\mathrm P_{3/2})$, if the molecule dissociates via the first-excited state, and $\mathrm I + \mathrm{Br}^*(^2\mathrm P_{1/2})$, if the dissociation takes place via the second-excited state. In the dissociation limit far away from the Franck-Condon region, the first-excited (spin-adiabatic) state acquires a $\mathrm Y(0^+)$ (spin-diabatic) character, whereas the second-excited (spin-adiabatic) state acquires a $\mathrm B(^3\Pi_{0^+})$ (spin-diabatic) character. Therefore, determining the populations of the electronic states at the end of the process allows us to determine the branching ratio of the dissociation products
\begin{align}\label{eqn: Q}
Q = \frac{[\mathrm I + \mathrm{Br}^*]}{[\mathrm I + \mathrm{Br}]+[\mathrm I + \mathrm{Br}^*]}
\end{align}
\begin{figure}
\centering
\includegraphics[width=.48\textwidth]{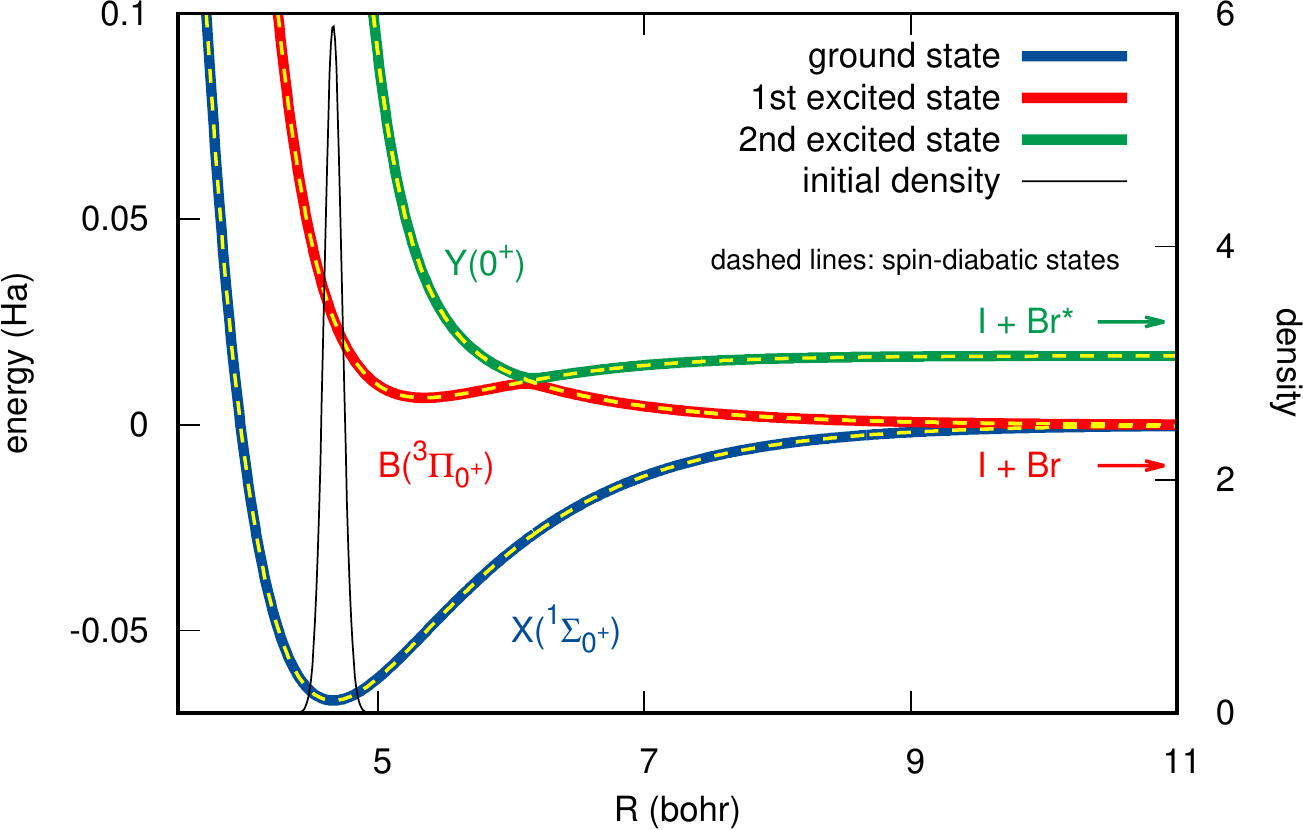}
\caption{Potential energy curves (PECs) corresponding to the spin-(a)diabatic states considered to describe the photo-dissociation process of IBr. Continuous colored lines are the spin-adiabatic PECs: ground state in blue, first excited state in red, second excited state in green. They are labeled using the notation of Ref.~\cite{Leone_Science2019} in the Franck-Condon region. Dashed yellow lines are spin-diabatic PECs. The initial nuclear density centered at the equilibrium position of the ground state PEC is shown as a thin black line. The arrows indicate the dissociation channels along the two electronic states.}
\label{fig: PES}
\end{figure}
In our simulations we use the model Hamiltonian of Ref.~\cite{Guo_JCP1993}, expressed in a spin-diabatic basis as
\begin{align}\label{eqn: el H}
\hat H_{el}(R) = \left(
\begin{array}{ccc}
H_0(R) & 0 & 0 \\
0 & H_1(R) & V_{12} \\
0 & V_{12} & H_2(R) 
\end{array}
\right)
\end{align}
where the nuclear coordinate $R$ is the IBr internuclear distance. The diagonal elements of the electronic Hamiltonian are the spin-diabatic potential energy curves (PECs), whereas the off-diagonal elements are the SOC, which are chosen to be constant. Following Ref.~\cite{Guo_JCP1993} we have
\begin{align}
H_0(R) &= A_0 \left[\left(1-e^{-\alpha_0(R-R_0)}\right)^2-1\right] \\
H_1(R) &= A_1\left[\left(1-e^{-\alpha_1(R-R_1)}\right)^2-1\right]+\Delta\\
H_2(R) &=  A_2\,e^{-\alpha_2\,R}+B_2\,e^{-\beta_2\,R}
\end{align}
where $A_0=0.067$~Ha, $\alpha_0=0.996$~bohr$^{-1}$, $R_0=4.666$~bohr, $A_1=0.01019$~Ha, $\alpha_1=1.271$~bohr$^{-1}$, $R_1=5.3479$~bohr, $\Delta=0.01679$~Ha, $A_2=2.82$~Ha, $\alpha_2=0.9186$~bohr$^{-1}$, $B_2=3.0\times 10^{7}$~Ha, $\beta_2=4.3$~bohr$^{-1}$, and $V_{12}=0.0006834$~Ha.

The spin-diabatic PECs are shown in Fig.~\ref{fig: PES} as yellow dashed lines. The curves corresponding to $H_1(R)$ and $H_2(R)$ cross at $R_c=6.2$~bohr, and in fact, after diagonalization of the Hamiltonian in Eq.~(\ref{eqn: el H}), an avoided crossing appears at $R_c$. The ground state is not coupled to the excited states, as clearly shown in the Hamiltonian of Eq.~(\ref{eqn: el H}) . However, as the transition dipole moment between $\mathrm X(^1\Sigma_{0^+})$ and $\mathrm B(^3\Pi_{0^+})$ is strong, an optical transition can be induced via an external field.

In this studied case, we suppose an instantaneous initial excitation. The molecule is prepared in the ground state, and the initial density is shown in Fig.~\ref{fig: PES}. Since the ground state PEC is nearly harmonic around the equilibrium position at $R_0=4.666$~bohr, the initial nuclear wavefunction $\chi(R,t=0)$ is a (real) Gaussian function
\begin{align}\label{eqn: init n wf}
\chi(R,t=0) = \sqrt[4]{\frac{1}{\pi\sigma^2}}e^{-\frac{(R-R_0)^2}{2\sigma^2}}
\end{align}
with variance $\sigma=0.096$~bohr. The nuclear mass used in the calculations is the IBr reduced mass $M=90023$~amu. The instantaneous excitation promotes the ground state wavepacket to the first excited state without geometric rearrangements, thus the first excited state is fully populated at time $t=0$, and all along the dynamics, the ground state remains non-populated because it is not coupled to the excited states.

CT-MQC dynamics is performed in the spin-adiabatic and in the spin-diabatic basis considering explicitly the 3 electronic states. Initial conditions are randomly sampled in position-momentum space from the (Gaussian) Wigner distribution associated to the initial nuclear state of Eq.~(\ref{eqn: init n wf}). For all calculations, we run $N_{tr}=1000$ coupled trajectories. For all trajectoires, the electronic coefficients corresponding to the initially populated state are set equal to one.

\begin{figure}
\centering
\includegraphics[width=.48\textwidth]{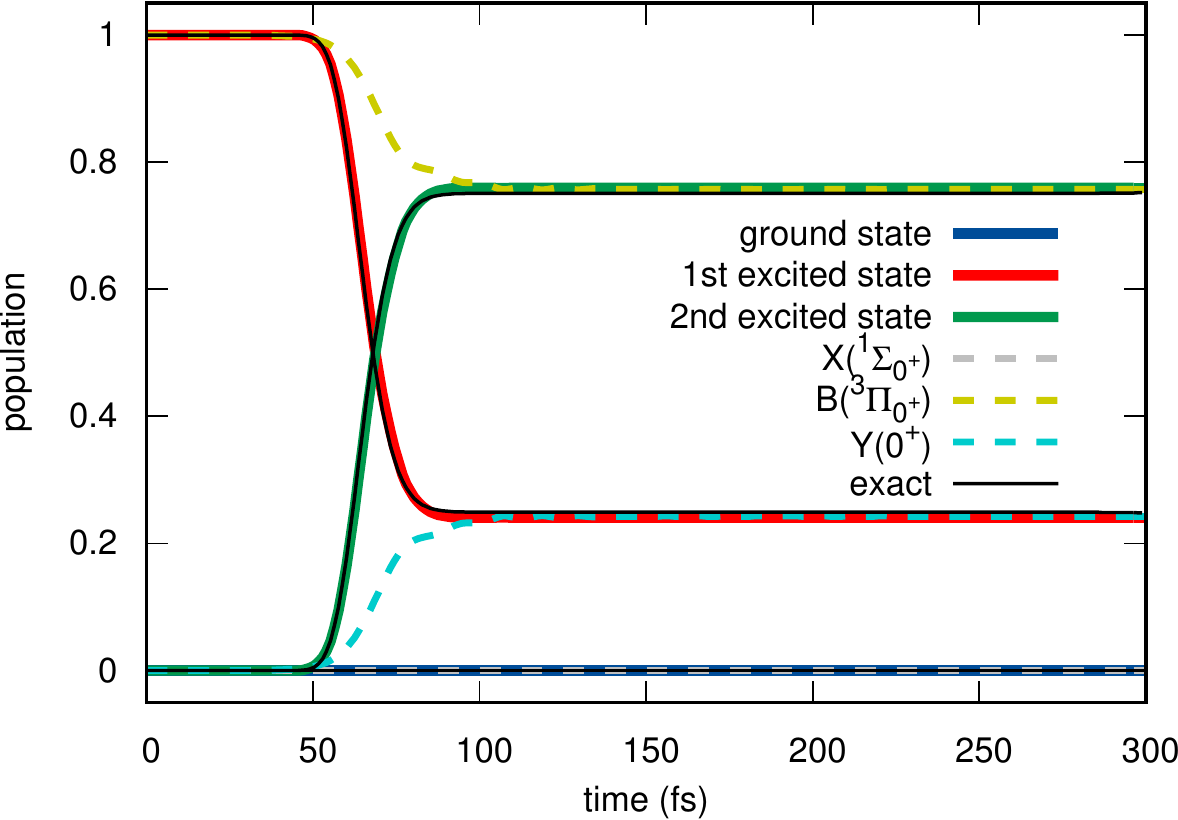}
\caption{Populations of the spin-(a)diabatic states as functions of time. Continuous lines are used to indicate spin-adiabatic populations: ground state in blue, first excited state in red, second excited state in green. Dashed lines are used to indicate spin-diabatic populations: $\mathrm X(^1\Sigma_{0^+})$ in gray, $\mathrm B(^3\Pi_{0^+})$ in dark-yellow, $\mathrm Y(0^+)$ in cyan. Exact spin-adiabatic results are shown as thin black lines.}
\label{fig: pop condon}
\end{figure}

When the molecule reaches $R_c$, population is partially transferred from the first excited state to the second excited state, and transfer is concluded just after 100~fs. In Fig.~\ref{fig: pop condon}, the populations of the spin-adiabatic states are shown as continuous lines as functions of time, whereas dashed lines are used to indicate spin-diabatic populations.
Comparison with exact quantum dynamics (thin black lines) shows that CT-MQC results reproduce extremely well the spin-adiabatic populations. As shown in Fig.~\ref{fig: PES}, the spin-diabatic PECs are identical with the spin-adiabatic curves in the asymptotic regions, but they differ in the vicinity of $R_c$. Therefore, between 50~fs and 100~fs, when the population transfer process takes place in Fig.~\ref{fig: pop condon}, the behavior of the populations differs. However, at long time, the spin-diabatic and spin-adiabatic populations agree: the population of the first excited state (red line in Fig.~\ref{fig: pop condon}) is the same as the population of the state $\mathrm Y(0^+)$ in the dissociating limit (dashed cyan line in Fig.~\ref{fig: pop condon}); the population of the second excited state (green line in Fig.~\ref{fig: pop condon}) is the same as the population of the state $\mathrm B(^3\Pi_{0^+})$ in the dissociating limit  (dashed dark-yellow line in Fig.~\ref{fig: pop condon}). Therefore, the branching ratio of the dissociation products can be determined by using either set of results; numerical values are reported in Table~\ref{tab: Q}.
\begin{table}[h!]
\centering
\begin{tabular}{c||c|c|c}
 & \phantom{a}exact \phantom{a}& spin-adiabatic& spin-diabatic\\
\hline
\hline
\phantom{a}$Q$\phantom{a}& \textbf{75~\%} & 76(1)~\% & 76(1)~\%  \\
%\hline
%Gaussian pulse & \textbf{75~\%}  & 83(1)~\%  & 80(1)~\%  
\end{tabular}
\caption{Branching ratio of the products of the photo-dissociation reaction of IBr. The reference values (exact) are determined based on quantum mechanical simulations using Eq.~(\ref{eqn: Q}). The branching ratio is also calculated based on CT-MQC simulations in the spin-adiabatic and spin-diabatic basis.}
\label{tab: Q}
\end{table}

In Table~\ref{tab: Q} the branching ratios calculated based on CT-MQC results are compared to the exact values calculated from the results shown in Fig.~\ref{fig: pop condon}. For CT-MQC, the errors on the branching ratios (in parenthesis) are determined from the standard deviations on the predictions of the final populations, and, within the error, CT-MQC values are in perfect agreement with the reference.

Finally, we show the comparison between nuclear quantum wavepacket dynamics and the trajectories.
\begin{figure}
\centering
\includegraphics[width=.48\textwidth]{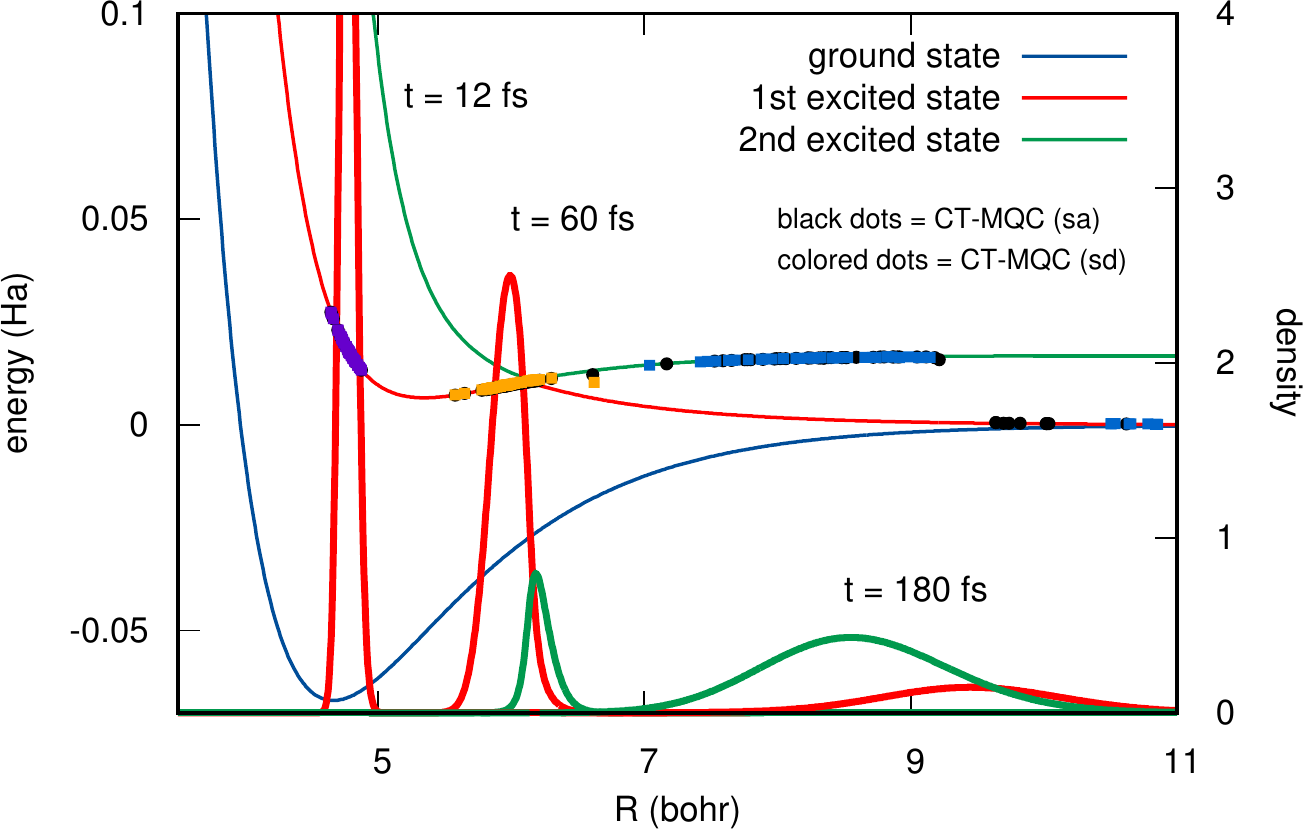}
\caption{PECs corresponding to the spin-adiabatic states are shown as thin colored lines. The nuclear densities corresponding to the first excited state (thick red lines) and to the second excited state (thick green lines) are shown at three time steps: $t = 12, 60, 180$~fs, as indicated in the figure. At the same times, the distributions of CT-MQC trajectories evolved in the spin-adiabatic (black dots) and in the spin-diabatic (colored dots) basis are shown. In particular, we use the value of the TDPES from Eq.~(\ref{eqn: TDPES}) at the position of each trajectory.}
\label{fig: traj}
\end{figure}
In Fig.~\ref{fig: traj}, the distributions of CT-MQC trajectories along the PECs that guide their evolution is superimposed to the nuclear wavepackets corresponding to the first excited state (red) and to the second excited state (green). Three snapshots at the time steps indicated in the figure are shown. CT-MQC equations are solved in the spin-adiabatic (black dots) and in the spin-diabatic (colored dots) basis, thus two sets of classical results are shown at the selected time steps. A very good agreement between the two sets of CT-MQC results, as well as with exact results, is observed.

In general, based on the above observations and on previous work, we conclude that CT-MQC is a reliable and flexible algorithm for trajectory-based calculations of light-induced ultrafast phenomena, involving nonadiabatic and spin-orbit coupling. Additional developments can be envisaged, especially aiming to refine the way the coupling of the trajectories is treated in the calculation of the quantum momentum, and to implement the effect of an external time-dependent field beyond the Floquet formalism.

\section{State of the art and perspectives}\label{sec: perspectives}
In the past few years, the exact factorization has been developed in different \textsl{flavors} to describe various kinds of systems and processes.

In its original time-dependent electron-nuclear formulation proposed by Gross and co-workers~\cite{Gross_PRL2010}, studies addressed the question as to what forces drive nuclear dynamics in nonadiabatic regimes~\cite{Gross_PRL2013,Gross_JCP2015,Agostini_JPCL2017}, ultimately leading to the derivation of CT-MQC~\cite{Gross_PRL2015}. Those studies focused on the characterization of the time-dependent potential energy surface and of the time-dependent vector potential of the theory in key situations manifesting strong nonadiabaticity with decoherence~\cite{Gross_PRL2013}, quantum interferences~\cite{Curchod_JCP2016}, and conical intersections~\cite{Agostini_JPCL2017}. In a similar spirit, and inverting the role of electronic and nuclear coordinates, the \textsl{inverse factorization} has been proposed to analyze the time-dependent Schr\"odinger equation for electronic dynamics with non-classical nuclei~\cite{Suzuki_PRA2014}. In situations of weak nonadiabaticity, instead, for instance when the nuclei move slower than the electrons, the electronic equation can be solved perturbatively~\cite{AgostiniEich_JCP2016}, the Born-Op\-pen\-hei\-mer regime being the unperturbed state of the system. This idea has been successfully applied to compute electronic flux densities ``within'' the Born-Oppenheimer approximation~\cite{Schild_JPCA2016}, the response of chiral molecules to infrared left and right circularly polarized light, known as vibrational circular dichroism~\cite{Scherrer_JCP2015}, and to estimate nonadiabatic corrections to infrared spectra of hydrogen-based molecules~\cite{Scherrer_PRX2017}.

As discussed previously, the exact factorization naturally lends itself to the inclusion of (classical) time-de\-pen\-dent external fields, for instance, laser pulses or cw lasers, to describe phenomena such as photo-induced ultrafast dynamics~\cite{Gross_JPCL2017}, strong field ionization~\cite{Maitra_PCCP2017,Maitra_PRL2015}, or periodically driven processes~\cite{Schmidt_PRA2017}. This is because the formalism already accounts for time-dependent potentials to describe the coupling between electrons and nuclei. In fact, the time-dependent potentials are modified by the presence of the external field, and have been analyzed to unravel dynamical details of dissociation~\cite{Gross_JCP2012}, electron localization~\cite{Suzuki_PRA2014,Suzuki_PCCP2015}, charge-resonance enhanced ionization~\cite{Maitra_PRL2015} in H$_2^+$, or to challenge single-electron pictures~\cite{Schild_arXiv2021} to describe molecules in strong lasers~\cite{Gross_PRL2017,Schild_PRR2020}. Recently, the exact factorization and CT-MQC (F-CT-MQC algorithm) have been combined with the Floquet formalism to interpret laser-driven dynamics in terms of single- or multi-photon absorption and emission processes~\cite{Agostini_JCP2021_2}. 

Interesting work has been conducted in the framework of the stationary Schr\"odinger equation~\cite{Gross_PTRSA2014,Hunter_IJQC1975_1,Cederbaum_JCP2013}, by proposing the static, time-independent version of the exact factorization of the electron-nuclear wavefunction. In this context it is worth mentioning three major contributions. One is the study of the behavior of the scalar potential and of the vector potential in the presence of conical intersections and in relation to geometric and topological phases~\cite{Min_PRL2014,Requist_PRA2015,Requist_PRA2017}. This work is oriented towards finding an answer to the question: Is the appearance of the molecular geometric phase effects within the Born-Oppenheimer approximation an artefact? The traditional molecular Berry phase is intimately connected to the requirement of infinitely slow nuclear motion. While this notion of adiabatic transport is a beautiful mathematical concept, real-world nuclei move the way move (i.e., not adiabatically) and this makes it difficult to relate the traditional molecular Berry phase to actual physical observables. By contrast, the geometric phase associated with the exact vector potential does not require any adiabaticity and, hence, it might be measurable. Even though conclusions of general validity are perhaps difficult to draw, enlightening case studies have been reported. Another topic that has been approached recently, and has been already applied to LiF, is the combination of the exact factorization formalism with density functional theory to solve the electronic equation, which is coupled to the nuclear Schr\"odinger equation~\cite{Gross_PRL2016,Gross_JCP2018}. Finally, a quantum electronic embedding method has been derived from the exact factorization in order to calculate static properties of a many-electron system~\cite{Gross_arXiv2019,Maitra_PRL2020}.

Motivated by experimental and theoretical advances of the past few years in the domain of physics and chemistry in cavities, the exact factorization of the electron-photon wavefunction and the exact factorization of the electron-nuclear-photon wavefunction have been proposed. Currently, studies are conducted focusing on the analysis of the properties of the scalar potential acting on the electrons as effect of the photons~\cite{Tokatly_EPJB2018} (reminiscent of the inverse factorization), as well as on the dynamic effect of the cavity on the electron-nuclear time-dependent potential energy surface~\cite{Maitra_EPJB2018,Maitra_PRL2019}, in analogy with polaritonic and cavity-Born-Oppenheimer surfaces. Application of the exact factorization in this domain is still preliminary, but the success of the original theory in many diverse fields strongly suggest that insightful novel developments are to be expected.

In the field of quantum molecular dynamics simulations, so far CT-MQC has proven its strength and flexibility. CT-MQC is a trajectory-based method, thus it is suitable for on-the-fly molecular dynamics calculations where electronic structure properties are determined along the dynamics only at the visited nuclear geometries~\cite{Gross_TDDFTbook2018}. The necessary electronic structure information to solve CT-MQC equations are energies, gradients and derivative couplings, along with spin-orbit coupling, when working in the spin-diabatic basis, and with the transition dipole moment, if the external time-dependent electric field is explicitly considered (beyond the Condon approximation). This information is available in standard quantum chemistry packages, thus CT-MQC can be combined with different approaches to electronic structure theory. It has been shown, in fact, that this is a viable route, employing time-dependent density functional theory and the CPMD software~\cite{CPMD}. The coupled nature of CT-MQC trajectories, however, remains the bottleneck even when parallellization strategies are employed, as was done in the CPMD implementation. To circumvent this issue, recently~\cite{Min_JCTC2018,Min_PCCP2019,Min_MP2019,Min_JPCL2018} the surface-hopping algorithm has been combined with the exact factorization including decoherence effects while maintaining (as much as possible) an independent-trajectory perspective. Further studies are ongoing to propose an algorithm/implementation able to access systems of ``experimental'' complexity. Interesting new developments are clearly envisaged, focusing on the theoretical formulation of CT-MQC, for instance aiming to capture nuclear quantum effects based on the quantum trajectory formulation of  nuclear dynamics, on its applications, addressing systems of growing complexity, and on the algorithmic aspect, with the purpose to improve the computational performance of such a coupled-trajectory-based scheme. 

\section{Conclusions} \label{sec: conclusions}
We reported an overview of the state of the art and perspectives on the application of the exact factorization of the electron-nuclear wavefunction in the field of quantum molecular dynamics.

We showed that the CT-MQC scheme is a reliable and flexible numerical procedure to solve the exact factorization equations with the support of classical-like trajectories. CT-MQC lends itself to on-the-fly ab initio molecular dynamics calculations to simulate and interpret highly nonequilibrium processes governed by various effects. Here, we presented a general method to treat nonadiabatic ultrafast dynamics, with spin-orbit effects, all within the same formalism.

Owing to the diverse achievements obtained so far and documented here, we envisage interesting new results of the ongoing developments.

%\bibliography{factorization,others}

\end{document}